\documentclass[
    aps,
    prd,
    reprint,
    superscriptaddress,
    nofootinbib,
    amsmath,
    amssymb,
    floatfix,
]{revtex4-2}

\usepackage{graphicx}
\usepackage{dcolumn}
\usepackage{bm}
\usepackage{hyperref}

\usepackage{physics}
\usepackage{xcolor}

\usepackage{siunitx} 

\newcommand{\SNR}{\mathrm{SNR}}

\begin{document}

\title{Einstein Telescope: the triangle in a network and the impact of the relative orientation of detectors}

\author{Giacomo Bruno}
\email{giacomo.bruno@uclouvain.be}
\affiliation{University of Louvain, Louvain-la-Neuve, B-1348, Belgium}

\author{Antonin Oswald}
\affiliation{University of Louvain, Louvain-la-Neuve, B-1348, Belgium}

\author{Federico De Lillo}
\affiliation{Universiteit Antwerpen, Prinsstraat 13, 2000 Antwerpen, Belgium}

\begin{abstract}
The geometrical configuration of the Einstein Telescope (ET) gravitational-wave observatory is one the most important design choices to be made for this future scientific facility. Two baseline configurations are currently being considered, a single triangle or 2 distant L-shaped detectors tilted by about $45^{\circ}$.  
In this document we present a study of the scientific performance of the ET baseline configurations in both stand-alone mode and in the presence of either an L-shaped Cosmic Explorer detector with 40-km long arms located in the United States or the existing LIGO-Virgo-KAGRA network. We also include in the study an alternative ET configuration, which is currently not being considered and consists of a triangle plus a distant L-shaped detector. 
We find that in stand-alone mode this alternative configuration performs significantly better than the other two in both the compact binary coalescence and stochastic gravitational wave background scientific cases.  
The differences in performance between the ET configurations are reduced in the presence of the considered, additional non-ET detectors. We also show that the impact of possible correlated noise in co-located detectors on the stochastic gravitational-wave background performance can be addressed with the addition of extra detectors.
Finally, we study the impact on the ET scientific performance of the relative orientation of the future detectors in the network. We find 
that, while the configurations that include a triangular ET detector are basically insensitive to this orientation, a balanced compromise for the ET configuration consisting of two L-shaped detectors could be represented by a tilt angle between the two detectors of about $30^{\circ}$.
\end{abstract}

\maketitle

\section{Introduction}
The Einstein Telescope (ET)~\cite{ET_Punturo:2010zza} is a future  European gravitational-wave (GW) observatory that is currently in its preparatory phase. ET will feature multiple underground detectors, each consisting of one cryogenic, lower laser-power interferometer optimised for detection in the low-frequency region (ET-LF, $\lesssim 20$ Hz) and one room-temperature, higher laser-power interferometer optimised for detection at higher frequencies (ET-HF, up to 10 kHz). The arms of the ET interferometers are expected to be at least 10 km in length, thus providing significant reduction of displacement noise compared to currently operating detectors, which have arms as long as 3 km (Virgo~\cite{aVirgo_VIRGO:2014yos} and KAGRA~\cite{kagra_KAGRA:2020tym}) or 4 km (LIGO~\cite{ALIGO_LIGOScientific:2014pky}). In the USA, another next-generation GW detector,
Cosmic Explorer~\cite{CE_Reitze:2019iox} (CE), is being prepared. CE will feature interferometer arms as long as 40 km, but will neither be cryogenic nor underground nor will operate at low laser power, which are features aimed to improve the GW sensitivity in the low-frequency region. If both ET and CE become realities, they will be able to operate jointly in a network of GW observatories, hopefully in the second half of the next decade. ET will also be able to operate jointly with the existing LIGO-Virgo-KAGRA (LVK) detectors and LIGO-India~\cite{Saleem:2021iwi}, which is a third LIGO detector under construction in India and featuring 4 km-long arms and sensitivity levels similar to the existing LIGO detectors.
Several studies have investigated in past years the science cases and configurations of a future network of GW detectors~\cite{Kalogera:2021bya, Evans:2021gyd, Evans:2023euw, COBA_Branchesi:2023mws, ET:2025xjr}, also including currently operating detectors~\cite{Gupta:2023lga}.

One of the most important parameter in the design of the ET detector is its geometrical configuration. Two options are currently being considered, a single triangle comprising three nested V-shaped detectors with an angle of $60^\circ$ between their arms, and 2 distant L-shaped detectors tilted by about $45^{\circ}$. In this document 
we also consider an alternative, hypothetical configuration consisting of a triangle plus a distant L-shaped detector, which we refer to as the 1T+1L option. This configuration is not currently under consideration.
We compare the performance of these three ET configurations with or without the presence of either an L-shaped CE detector with 40 km long arms located in the United States or the existing LVK network. 
We also study the impact on scientific performance of the relative orientation of the future GW detectors, ET and CE, in the various configurations.

The structure of this paper is as follows. We discuss  the investigated detector configurations in Sec.~\ref{sec:1T1Loption}.
We then go through the main ET science cases~\cite{ET:2025xjr} classified in terms of GW signal topology: compact binary coalescences (CBC), including multi-messenger astronomy, in Sec.~\ref{sec:CBC_study}, stochastic GW background (SGWB) in Sec.~\ref{sec:SGWB_study}, searches for unmodelled relatively short signals (``bursts'') in Sec.~\ref{sec:Bursts} and quasi-monochromatic long-duration signals (``continuous waves'', CWs) in Sec.~\ref{sec:CW}. For the first two cases we present results obtained in simplified setups. 
The last two science cases are simply briefly discussed.

\section{Detector configurations}
\label{sec:1T1Loption}
Both the single triangle, called 1T in the following, and the two distant L-shaped detectors tilted by $45^{\circ}$, 2L in short in the following~\footnote{whenever a different tilt angle between the two L-shaped detectors is considered, the tilt angle will be explicitly quoted}, provide exceptional performance in many of the scientific cases to be covered by ET, as demonstrated in~\cite{COBA_Branchesi:2023mws}. 
Nevertheless, as discussed below, each of them also displays some shortcomings that may significantly degrade the scientific reach of the ET observatory, especially in the absence of additional detectors. 

With six interferometers installed at a single site, the 1T option provides redundancy and
all the benefits of the null stream~\cite{Freise_2009, null_stream_Goncharov:2022dgl, null_stream_Janssens:2022cty}, which is a signal-free data stream obtained from the sum of the strain outputs of its three detectors.  The null stream facilitates the unbiased estimation of the detector noise power spectral density~\cite{PhysRevD.86.122001} even in the presence of correlated noise~\cite{null_stream_Janssens:2022cty} and glitches~\cite{null_stream_Goncharov:2022dgl}, which are short-duration transients in the detector output caused by instrumental or environmental disturbances. The null stream can also help cross-calibrate detector outputs~\cite{Schutz:2020hyz, Wong:2025iaf}, identify and reject glitches appearing in individual detectors~\cite{null_stream_Narola:2024qdh}, and discriminate genuine GW signals from instrumental effects~\cite{Chatterji:2006nh, PhysRevD.105.084002}. Nevertheless, a few points need to be noted. First, the availability of the null stream is conditioned to all three detectors being in observation mode~\cite{null_stream_Goncharov:2022dgl, Negr:2026gwp}. This condition is typically verified for the low-uncertainty CBC events considered in Sec.\ref{sec:CBC_study}, but may not be for a rare, unmodeled transient signal candidate for which discrimination from spurious noise coincidences is crucial (cf. Sec.\ref{sec:Bursts}). Second, the null stream independence of the source sky position and GW frequency is valid as long as the GW wavelength is large compared to the detector arm length~\cite{Schutz_2009}. For a triangular detector with 10 km-long arms this assumption starts to break down at GW frequency values greater than about 1 kHz~\cite{Virtuoso:2024kyp}. It should be noted that the impact of the null stream could not be taken into account in the results presented in either~\cite{COBA_Branchesi:2023mws} or in this document, because of the unavailability of fully realistic datasets that would include overlapping GW signals and non-Gaussian noise features.\\
At the same time the 1T option comes with a complex system of interferometers and is unable, in the absence of other distant detectors, to exploit the delay between the signal arrival times at the detectors for accurate sky localisation. Finally, the 1T option must also be considered in light of the uncertainty in the rate and nature of possible correlated noise of local environmental origin that may affect its co-located detectors. Examples of this kind of noise are seismic and Newtonian noise~\cite{Janssens:2022xmo} as well as magnetic noise\cite{Janssens:2021cta}. Numerous studies~\cite{Badaracco_2019, Badaracco_2024, Schillings_2025, Koley:2024wev,vanBeveren_2023, Ophardt:2025keu} have been addressing the question of Newtonian Noise mitigation with encouraging results.

On the other hand, the 2L option comes with fewer interferometers to build, commission, operate, and upgrade. The feasibility of longer 15 km arms, the built-in availability of a baseline for better sky localisation and the absence of possible correlated noise arising from local environmental sources are other attractive features of this option.
At the same time the 2L is a multi-site solution, thus featuring more complex logistics and the necessity of deploying laboratory personnel and detector experts over two sites. Moreover, the 2L misses all the benefits of the null stream mentioned above and, for relative orientations between the two detectors close to $45^\circ$, strongly penalises the case of SGWB searches (cf. Sec.~\ref{sec:SGWB_study}). 
More generally, as discussed in the following and in past works~\cite{COBA_Branchesi:2023mws, 3G_network_Ebersold:2024hgp, Begnoni:2025oyd, Iacovelli:2024mjy}, the 2 L-shaped detector option displays a stand-alone physics performance that in several cases significantly depends on the relative orientation of its two detectors and will possibly have to face constraints on this orientation arising from construction technical reasons.   

From the above discussion it is clear that the performance of the two ET baseline configurations, the 1T in particular, can be improved if additional detectors are included in a network. In the following we consider as additional detectors either the future CE or the existing LVK detectors with the sensitivity expected during the O5 observing run. The choice of LVK with O5 sensitivity as the additional supporting network composed of current-generation detectors is arbitrary and simply driven by the need to identify a realistic scenario in terms of overall performance of this supporting network at the start of ET operation. In reality, this final supporting network could have significantly higher performance in a scenario where all existing detectors continue operation in the CE and ET era together with LIGO-India. Moreover, assuming that common technological solutions will be adopted for ET-HF, CE, and the other observatories, the sensitivity of current-generation detectors could also be significantly higher than what assumed in this study. 
We also consider the alternative 1T+1L ET configuration mentioned above. We claim that this configuration can provide a robust and high-performance European infrastructure, capable of covering the full ET scientific program in stand-alone mode. Stand-alone performance is particularly important in the very low-frequency region ($\lesssim 10$ Hz) where it is reasonable to expect that, even in the presence of additional detectors, ET will be close to operating alone. Indeed, underground operation, low laser power and cryogenic mirrors should give a decisive edge to ET in this frequency region, where it should eventually display significantly reduced seismic, Newtonian, thermal and radiation pressure noise.\\
The 1T+1L option clearly comes at a cost, namely increased overall complexity and construction costs, a double-site infrastructure (as for the 2L option) and higher maintenance, operation and upgrade needs. From the operational point of view, however, there are strong indications that this configuration can be successful. 
First, the LIGO scientific Collaboration has demonstrated that it is possible to operate successfully a multi-site observatory. Second, a lot of effort has been going into the development of solutions aiming to automate commissioning and operating procedures for GW detectors~\cite{Biswas:2019wmx, Qin:2025kie, Ma:2023csv, Mukund:2023oyy}. This trend is expected to be continued, also in view of the fast development of machine-learning techniques~\cite{Cuoco:2024cdk}. 

\begin{figure}
\includegraphics[width = \columnwidth]{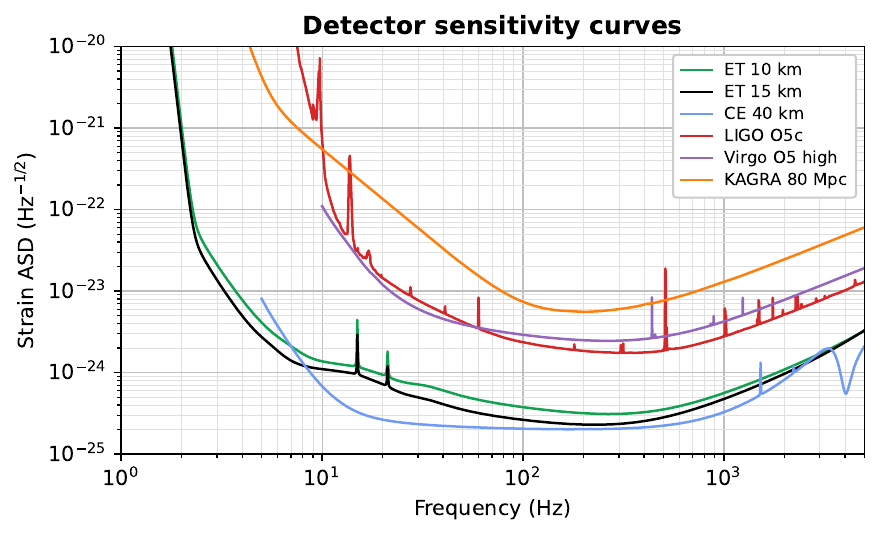}
\caption{Detector sensitivity curves adopted in this study and expressed as noise amplitude spectral density normalised to detector strain. The curves are presented for a single ET L-shaped detector with 10 km and 15 km long arms~\cite{COBASensRepository}, a CE L-shaped detector with 40 km long arms \cite{CESensRepository}, LIGO ``O5c''~\cite{ligo_projection}, Virgo ``O5 high-noise''~\cite{lvk_projection}, and KAGRA ``80 Mpc''~\cite{lvk_projection}.}
\label{fig:detsens}
\end{figure}

The three ET configurations investigated in this document are reported in Table~\ref{tab:detconf}.
Specifically, the 1T option has 10 km long arms and is placed in the Euregio Meuse-Rhin (EMR) region at the frontier between Belgium, Germany and the Netherlands, which is one of the three ET candidate sites. One side of the triangle is along the local West-East direction and the opposite vertex points toward the local North. 
The 2L configuration has detectors with 15 km-long arms. 
One of the two detectors is placed in Sardinia, which is another ET candidate site, while the second detector is placed in the EMR region.
As mentioned above, unless the relative orientation of the two detectors is specified otherwise, we will assume an angle of $45^\circ$, using local geographical coordinates as reference, and refer to this specific configuration in short as 2L. This configuration could also be realised in Sardinia and Saxony, which is the third ET candidate site, but the choice described above was made to align the results presented in this paper to those in~\cite{COBA_Branchesi:2023mws}. The results are not expected to change significantly if two different sites are chosen. 
The 1T+1L configuration consists of a 10 km long arms triangle in the EMR region and an L-shaped detector with 15 km long arms in Sardinia.  
The CE detector considered in this study consists of a single L-shaped detector with 40 km arms placed in Idaho, United States of America. 
Unless specified otherwise, the default positions of the ET and CE detectors are reported in Table~\ref{tab:detplac}.\\ 
The detector sensitivities used in this study are expressed in terms of noise amplitude spectral density normalised to detector strain and are reported in Fig.~\ref{fig:detsens}. They are obtained from~\cite{COBASensRepository} and~\cite{CESensRepository} for ET and CE, respectively. For the LVK detectors, the following projected sensitivities have been used: the ``O5c'' projection for LIGO, corresponding to the last of the three O5 epochs foreseen in~\cite{ligo_projection}, and the O5 projection for Virgo (``high-noise limit'') and KAGRA (``80 Mpc'') both taken from~\cite{lvk_projection}.

The locations and orientations of the 1T, 2L and CE detectors as well as the adopted sensitivity curves are the same that were used in~\cite{COBA_Branchesi:2023mws}, except for the fact that in~\cite{COBA_Branchesi:2023mws} the triangle was placed in Sardinia, whereas in this study it is placed in the EMR region.

\begin{table*}[htbp]
\centering
\begin{tabular}{lcc}
\hline
Detector & Arm length (km)  & Location \\
\hline
1T    & 10            & EMR \\
2L    & 15, 15        & EMR and Sardinia \\
1T+1L & 10, 15        & EMR and Sardinia \\
CE    & 40            & Idaho \\
\hline
\end{tabular}
\caption{Configurations of the future detectors investigated in this document. Unless specified otherwise, the relative orientation between the detectors in the double-L option
is $45^\circ$ (2L)}
\label{tab:detconf}
\end{table*}

\begin{table*}[htbp]
\centering
\begin{tabular}{lccc}
\hline
Detector & Latitude ($^\circ$) & Longitude ($^\circ$) & Default orientation  \\
\hline
T in EMR &
\ang{50;43;23} &
\ang{5;55;14} &
one side parallel to the local East-West direction. Opposite vertex pointing North. \\

L in EMR &
\ang{50;43;23} &
\ang{5;55;14} &
$x$-arm ($y$-arm) pointing to local North-East (North-West) \\

L in Sardinia &
\ang{40; 31; 0} &
\ang{9; 25; 0} &
$x$-arm ($y$-arm) pointing to local East (North) \\

CE &
\ang{43;49;36} &
\ang{-112;49;30} &
x-arm (y-arm) at $56^\circ$  counterclockwise from local North (West)\\
\hline
\end{tabular}
\caption{Location coordinates and default orientation of the future detectors investigated in this study.}
\label{tab:detplac}
\end{table*}


\section{CBC and multi-messenger astronomy}
\label{sec:CBC_study}

\subsection{Methods and data}

The detection of CBC events, where the system is composed of black holes (BHs) and/or binary neutron stars (NSs), and the accurate measurement of their parameters is fundamental for testing General Relativity, performing population studies, studying the internal dynamics of NSs or searching for GW lensing. In addition, the accurate measurements of the sky localisation and luminosity distance of CBC events will play a primary role in multi-messenger astronomy~\cite{Corsi:2024vvr} and stand-alone Hubble constant measurement~\cite{Borhanian:2020vyr}. A comprehensive review of the CBC science in the ET era can be found in~\cite{ET:2025xjr}. 

Several studies~\cite{Dupletsa:2022scg, Ronchini:2022gwk, Pieroni:2022bbh, Borhanian:2022czq, Gardner:2023znk, 3G_network_Maggiore:2024cwf, 3G_network_Ebersold:2024hgp, DeRenzis:2024dvx, Begnoni:2025oyd, Iacovelli:2024mjy, COBA_Branchesi:2023mws, Borhanian:2025uni, Santoliquido:2025aiq} have investigated in the past the CBC scientific performance of ET in both stand-alone mode and in an extended network.
The analysed dataset of CBC events and the software used for obtaining the results presented in this section are the same as those adopted in~\cite{COBA_Branchesi:2023mws}. Their description is briefly summarised below. 

The dataset of CBC events is publicly available~\cite{CBCCatalog} and represents the binary NS (BNS) and binary BH (BBH) coalescences that can be expected to be detectable in one year of operation of ET. All the details related to the generation of these events, including the adopted population and waveform models, can be found in~\cite{COBA_Branchesi:2023mws}.  

The \verb|GWFAST| software~\cite{Iacovelli:2022bbs,GWFAST_Iacovelli:2022mbg}, which is based on the Fisher matrix formalism, has been used for signal-to-noise ratio (SNR) calculation and parameter estimation. 
The waveform models \verb|IMRPhenomXPHM|~\cite{bbh_model} and \verb|IMRPhenomD_NRTidalv2|~\cite{bbh_model} are used for BBH and BNS events, respectively. 
\verb|GWFAST| takes into account the effect of the Earth's rotation on the detected signals, namely the amplitude and phase modulation caused by the variation of the antenna pattern functions~\cite{PhysRevD.58.063001} and a Doppler effect that results in the so-called location phase factor~\cite{Iacovelli:2022bbs}.

For both BBH and BNS events we adopt the same sets of waveform parameters that were used in~\cite{COBA_Branchesi:2023mws} and report results on 
the uncertainty in the estimate of the following ones:
the detector-frame chirp mass ($\mathcal{M}_c$), the symmetric mass ratio ($\eta$),  the luminosity distance to the source ($d_L$), the source sky position coordinates ($\theta$ and $\phi$), 
the polarisation angle ($\psi$), the angle between
the orbital angular momentum of the binary system and the line of sight ($\iota$), the angle between the total angular momentum and the line of sight ($\theta_{JN}$, which is the same as $\iota$ only in the absence of precession), the dimensionless spin of the compact object $i=1,2$ along the direction of the orbital angular momentum of the binary system ($\chi_i$) and the dimensionless tidal deformability ($\Lambda_i$, only for BNS events).
The parameters $\theta$ and $\phi$ are defined as $\theta =  \pi/2 - \delta$
and $\phi = \alpha$, with $\delta$ and $\alpha$ declination and right ascension, respectively.
The labels `1' and `2' refer, respectively, to the heaviest and lightest component of the binary system. Also, instead of $\Lambda_1$ and $\Lambda_2$, the two combinations 
$\tilde{\Lambda}$ and $\delta \tilde{\Lambda}$ defined in~\cite{Wade:2014vqa} are used. 

Statistical inference is performed on all waveform parameters. For BBHs, $\theta_{JN}$ and the magnitude and angles of the objects' spins are used in the place of $\iota$ and $\chi_i$. For BNSs, given the small expected values of their spin magnitudes, only the
spin components aligned to the system orbital angular momentum are used in the analysis.
More details on the applied inference procedure can be found in~\cite{Iacovelli:2022bbs}.

Following~\cite{COBA_Branchesi:2023mws}, each detector
is assumed to have a duty cycle of 85\% that is uncorrelated with that of the other detectors. The starting frequency of the analysis is 2 Hz. Both BBH and BNS events are required to satisfy SNR $\geq 12$, and have Fisher matrix inversion uncertainty smaller than $5 \times 10^{-2}$~\cite{Iacovelli:2022bbs}. The latter requirement discards less than 1\% and 6\% of the BBH and BNS events, respectively. These events are in most cases observed with a single detector, as a result of the duty cycle effect, and therefore have little scientific relevance because of the low resolution on the measurement of the system parameters. The reader is referred to \cite[footnote 9]{COBA_Branchesi:2023mws} for a more complete discussion on this requirement.

\subsection{Results}

\begin{figure*}
    \centering
    \includegraphics[width=\linewidth]{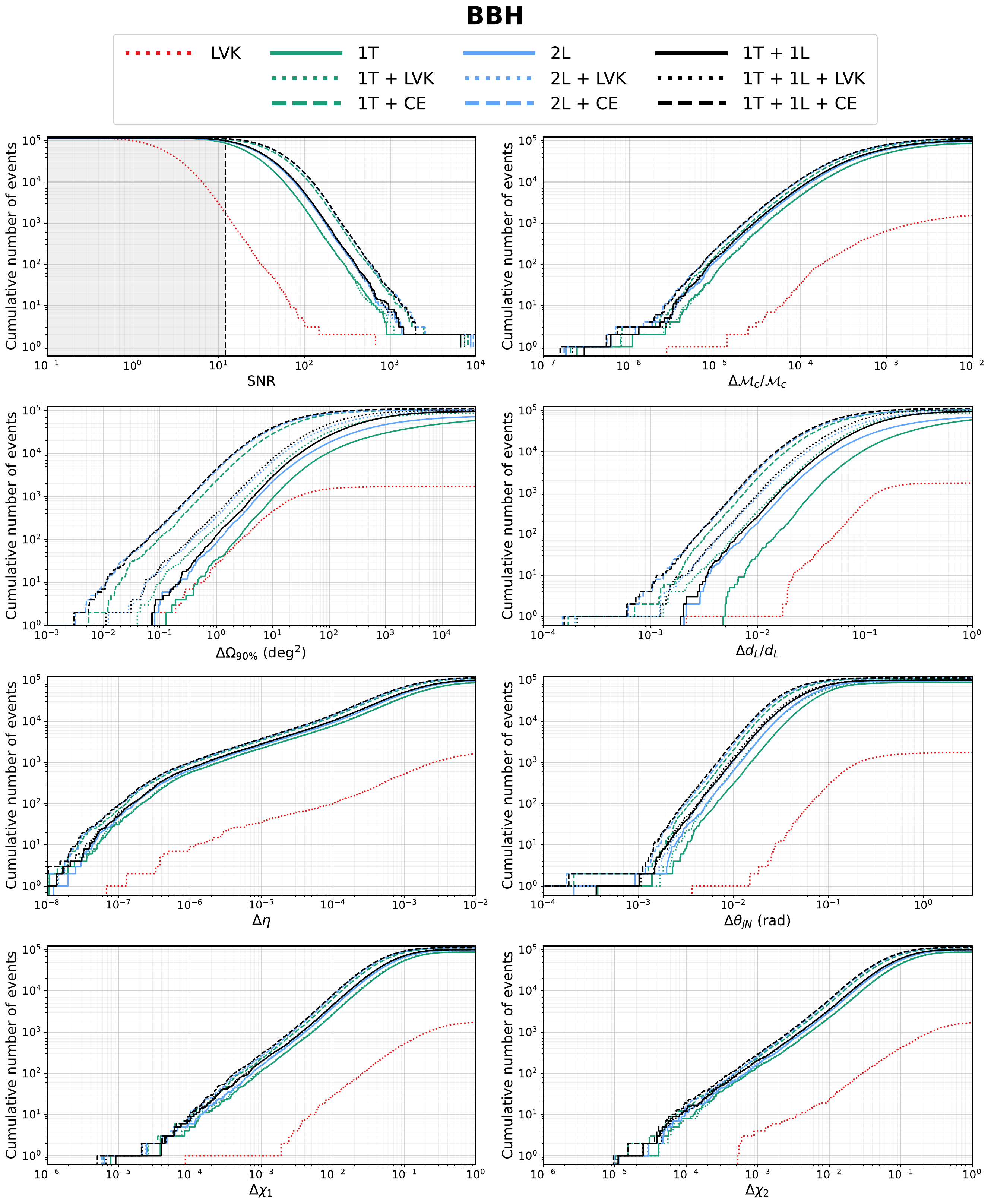}
    \caption{Cumulative number of BBH events detected per year as a function of SNR (top left plot) or parameter uncertainty (other plots) for the different ET configurations, with and without LVK or CE. The definition and calculation of the parameters as well as the generation and selection of the BBH events are discussed in the text.}
    \label{fig:bbh}
\end{figure*}

\begin{figure*}
    \centering
    \includegraphics[width=\linewidth]{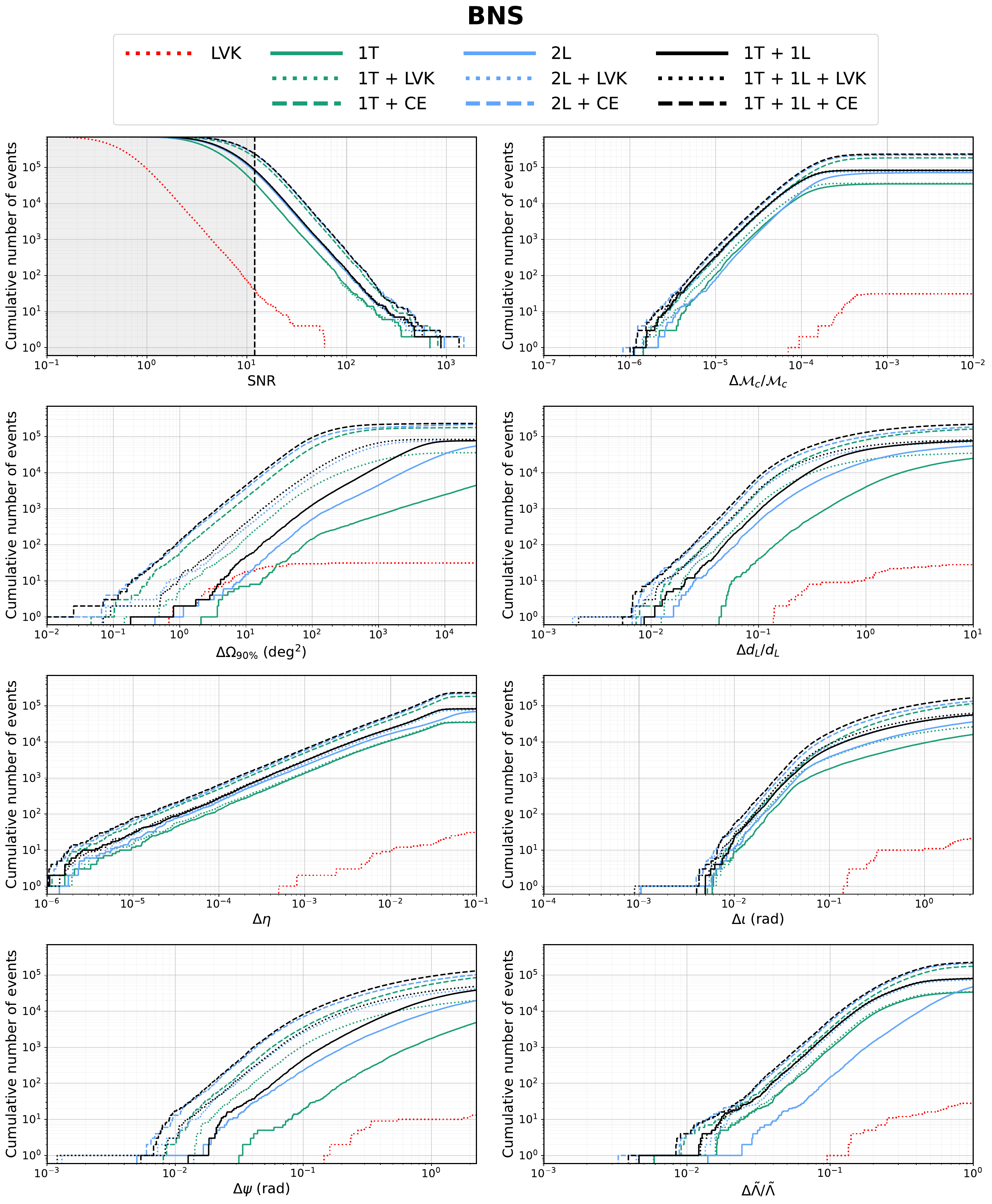}
    \caption{Cumulative number of BNS events detected per year as a function of SNR (top left plot) or parameter uncertainty (other plots) for the different ET configurations, with and without LVK or CE. The definition and calculation of the parameters as well as the generation and selection of the BNS events are discussed in the text.} 
    \label{fig:bns}
\end{figure*}

The cumulative number of BBH and  BNS events detected per year as a function of SNR or parameter uncertainty are reported in Figs.~\ref{fig:bbh} and ~\ref{fig:bns}, respectively, for the three different ET configurations, with and without LVK or CE~\footnote{It should be noted that in these figures the regions corresponding to the lowest achievable parameter uncertainty, which are also the most interesting from a scientific point of view, are in general affected by larger statistical uncertainties associated with the limited size of the data sample.}.
The parameter uncertainty is expressed in all cases as one standard deviation except for the sky localisation, for which the 90\% confidence level (C.L.) area around the fitted $\theta$ and $\phi$ central values is reported.

In terms of SNR the 1T+1L performs only slightly better than 2L, while these two configurations detect a factor of 2 to 3 more events than 1T at a given SNR threshold. The largest differences between the three ET configurations are observed in the uncertainty in the sky localisation and luminosity distance measurements, which are particularly important for their impact on multi-messenger astronomy and stand-alone Hubble constant measurement~\cite{Borhanian:2020vyr}, as mentioned earlier. For these two figures of merit the 1T+1L performs at least 10 times better than 1T and 2 to 3 times better than 2L. These differences become smaller for BBH events: 3 to 10 times better than 1T and 30\% to 50\% better than 2L, depending on the considered parameter and the threshold on its uncertainty. In the absence of additional detectors, the lack of a baseline allowing an accurate time delay measurement between detectors penalises the sky localisation with 1T. Instead, 1T+1L benefits, like 2L, from such a time delay measurement, and in addition exploits multiple signal measurements modulated during the duration of the signal by the effects mentioned above associated with the time variation of the antenna pattern functions of the detectors and the Doppler effect caused by the rotation of the Earth. The smaller differences in performance betweeen 1T+1L and 2L for BBH events compared to BNS ones are explained with the much shorter observable duration of the BBH events, implying a larger impact of the time delay measurement. \\

The performance of the three ET configurations is highly improved with the addition of CE, which also reduces significantly the differences in performance between the 3 configurations\footnote{This conclusion was already reached in~\cite{COBA_Branchesi:2023mws, Evans:2021gyd, 3G_network_Maggiore:2024cwf, Iacovelli:2024mjy} for the 1T, 2L and two aligned L-shaped detectors. Notably, one of these studies~\cite{3G_network_Maggiore:2024cwf} also considered a network composed of a single L-shaped ET detector, a single CE detector and LIGO-India, finding competitive CBC performance.}. 
Nevertheless 1T+1L and 2L with CE still perform, in terms of sky localisation and luminosity distance uncertainty, about a factor of 2 better than 1T with CE.
Similarly, the addition of LVK improves significantly the ET performance in terms of sky localisation and luminosity distance uncertainty (and some other parameters) and also reduces the differences between the ET configurations, but not to the same extent as the addition of CE.

As partly anticipated in Sec.~\ref{sec:1T1Loption}, a full parameter estimation analysis capable of capturing possible multi modalities, accurately handling correlations among parameters in the posterior distributions, and performed on realistic ET data featuring overlap of multiple GW signals with both Gaussian and non-Gaussian noise, is not incorporated in either these results or in those presented in~\cite{COBA_Branchesi:2023mws}. Several past works have investigated the limitations of the Fisher matrix formalism~\cite{Vallisneri:2007ev, Iacovelli:2022bbs, Dupletsa:2024gfl} and have compared it with full Bayesian methods~\cite{Santoliquido:2025lot, Santoliquido:2025aiq}. Moreover, the availability of the null stream in the 1T and 1T+1L may provide significant advantages to these two options for the reasons discussed in Sec.~\ref{sec:1T1Loption}. Therefore the quantitative conclusions that have been drawn in this section may likely change in such more realistic conditions. Dedicated studies, which go beyond the scope of this document, should be conducted to shed light on these matters.

\begin{figure*}
    \centering
\includegraphics[width=\linewidth]{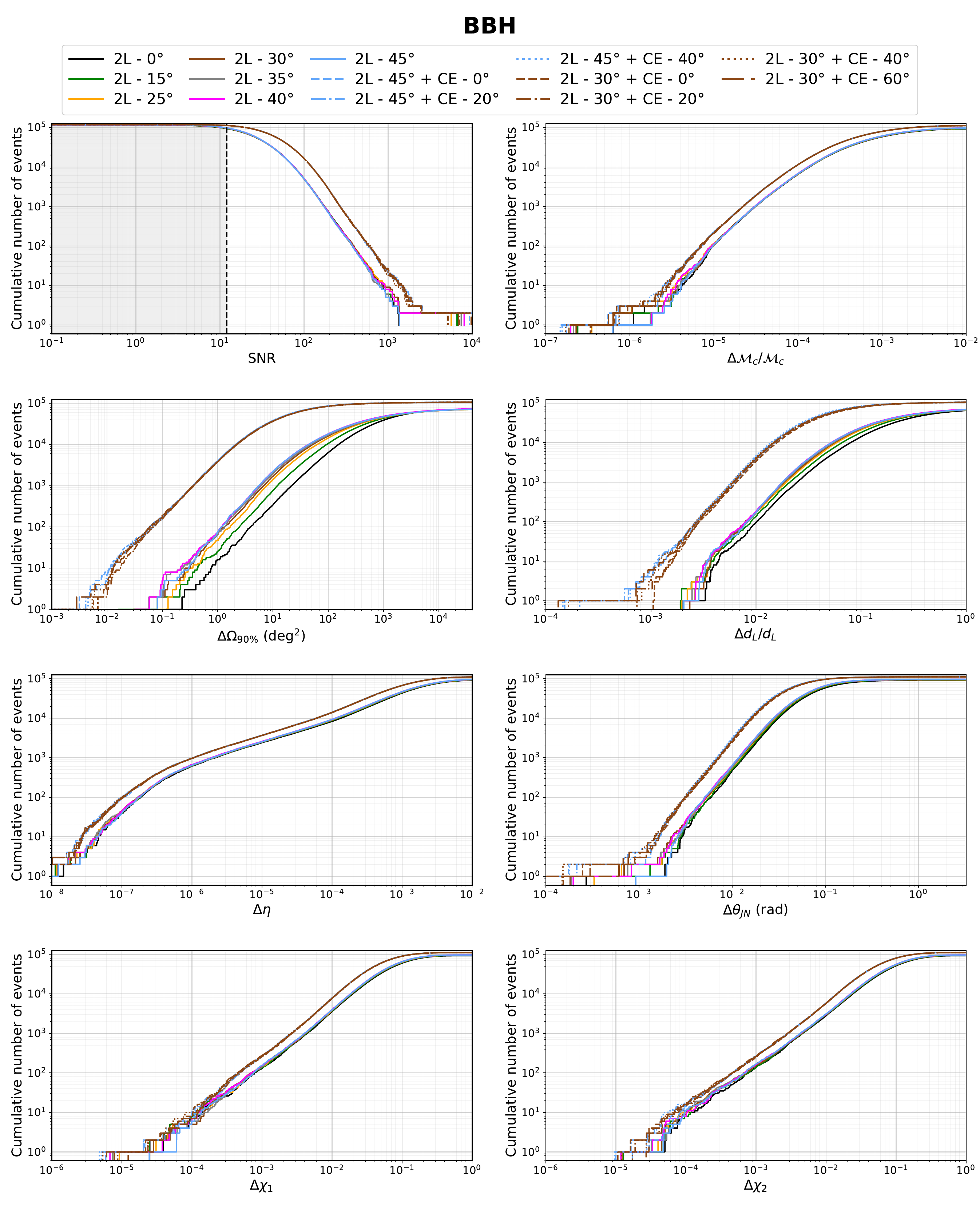}
    \caption{Cumulative number of BBH events detected per year as a function of SNR (top left plot) or parameter uncertainty (other plots) for various relative orientations of the two L-shaped  detectors in the ET double-L configuration, with and without CE. The angle by which the L-shaped detector in Sardinia is rotated clockwise with respect to the one in the EMR region, which is in its default position, is indicated in the legend next to the label ``2L-''. The CE detector is rotated clockwise, starting from its default position, by the angle indicated in the legend next to the label ``CE-''. The definition and calculation of the parameters as well as the generation and selection of the BBH events are discussed in the text.}
\label{fig:bbh_orientation}
\end{figure*}

\begin{figure*}
    \centering
    \includegraphics[width=\linewidth]{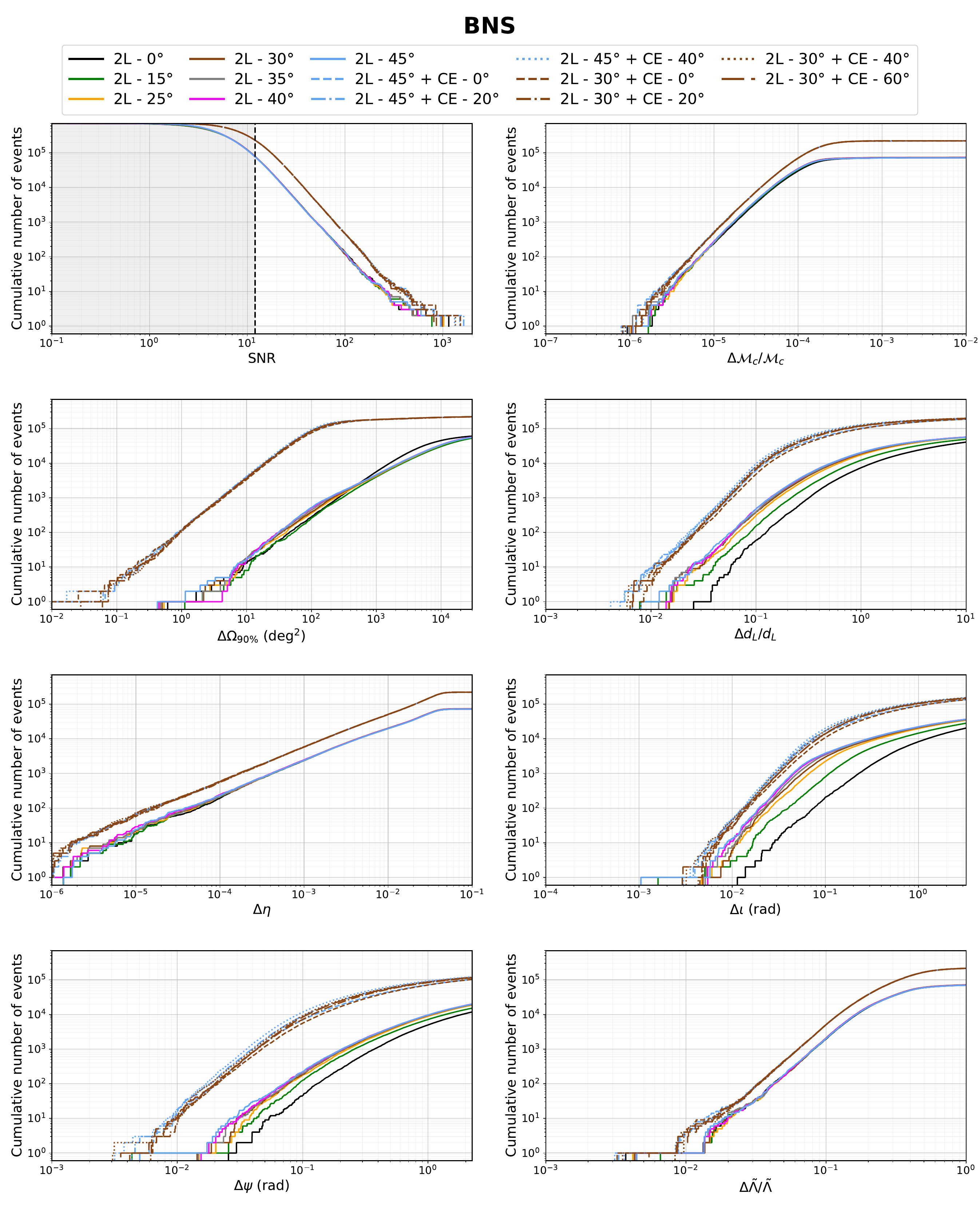}
    \caption{Cumulative number of BNS events detected per year as a function of SNR (top left plot) or parameter uncertainty (other plots) for various relative orientations of the two L-shaped  detectors in the ET double-L configuration, with and without CE. The angle by which the L-shaped detector in Sardinia is rotated clockwise with respect to the one in the EMR region, which is in its default position, is indicated in the legend next to the label ``2L-''. The CE detector is rotated clockwise, starting from its default position, by the angle indicated in the legend next to the label ``CE-''. The definition and calculation of the parameters as well as the generation and selection of the BNS events are discussed in the text.}
        \label{fig:bns_orientation}
\end{figure*}

\begin{figure*}[h]
    \centering
    \includegraphics[width=\linewidth]{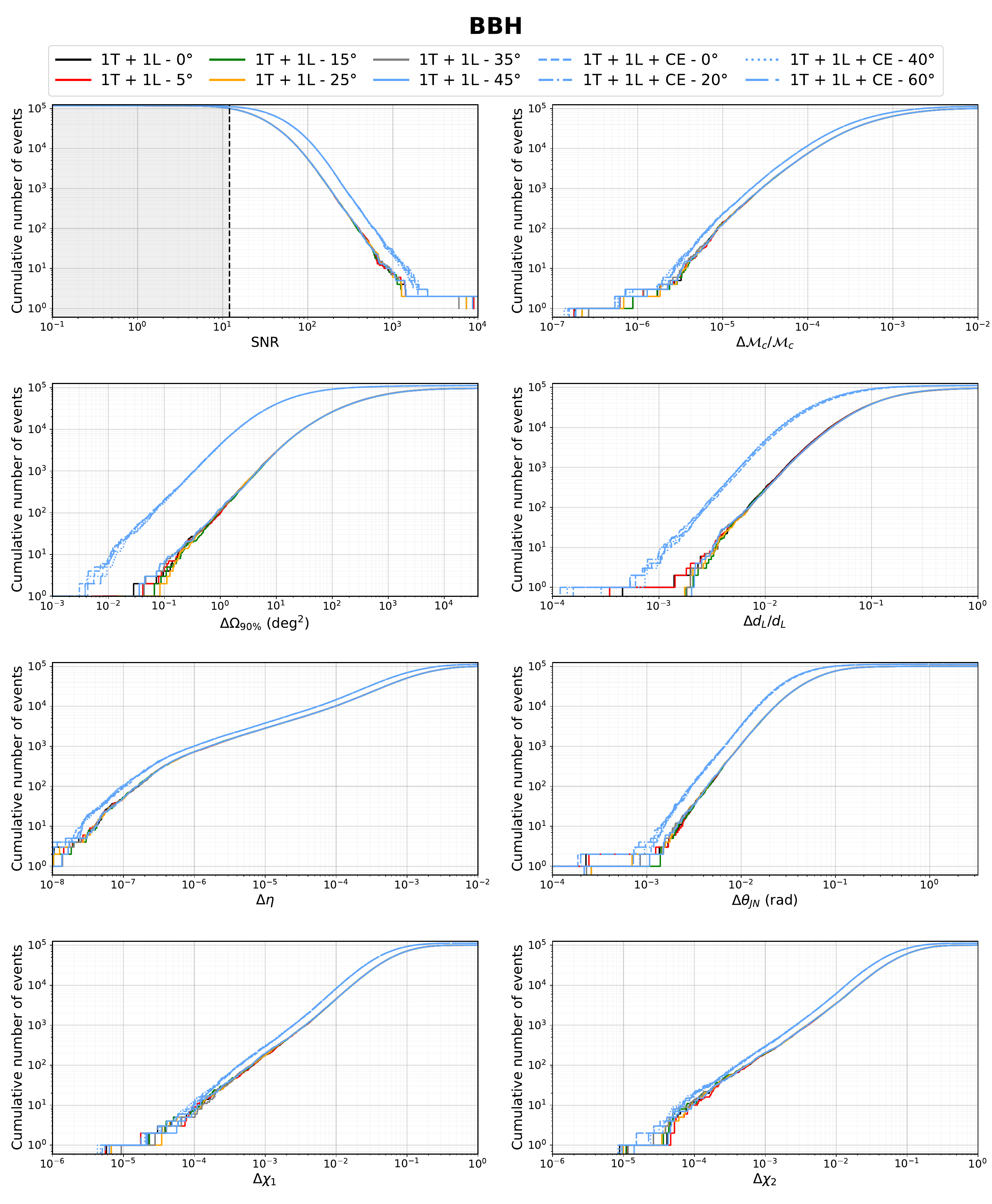}
    \caption{Cumulative number of BBH events detected per year as a function of SNR (top left plot) or parameter uncertainty (other plots) for various relative orientations of the L-shaped and triangular detectors in the 1T+1L ET configuration, with and without CE. The L-shaped detector in Sardinia is rotated clockwise, starting from the position where its x-arm (y-arm) points North-East (North-West), by the angle indicated in the legend next to the label ``1T+1L-''. The triangular detector in the EMR region is always in its default position. The CE detector is rotated clockwise, starting from its default position, by the angle indicated in the legend next to the label ``CE-''. The definition and calculation of the parameters as well as the generation and selection of the BBH events are discussed in the text.}
    \label{figCBCOrientationStudy_1T1LCE_bbh}
\end{figure*}

\begin{figure*}[h]
    \centering
    \includegraphics[width=\linewidth]{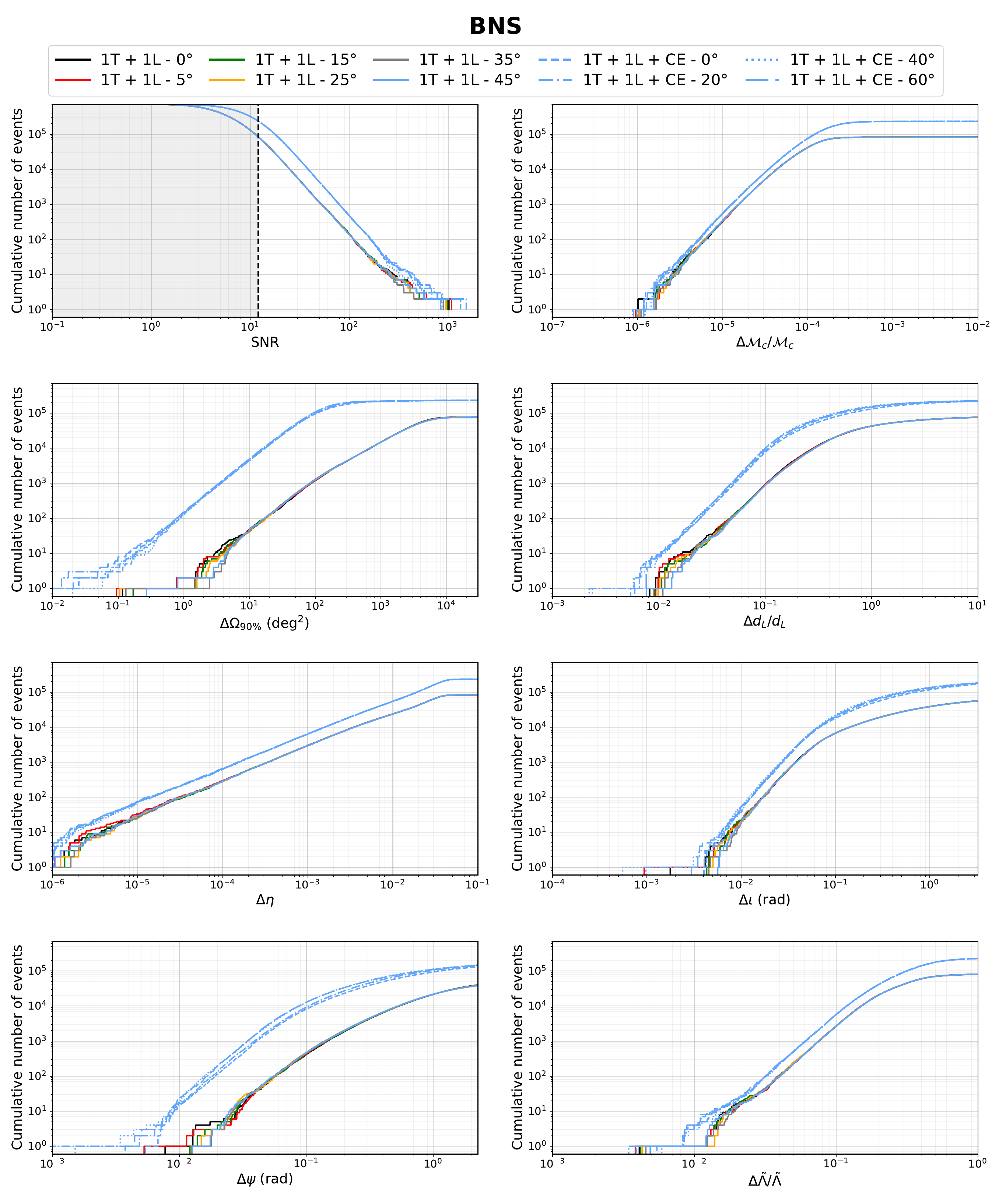}
    \caption{Cumulative number of BNS events detected per year as a function of SNR (top left plot) or parameter uncertainty (other plots) for various relative orientations of the L-shaped and triangular detectors in the 1T+1L ET configuration, with and without CE. The L-shaped detector in Sardinia is rotated clockwise, starting from the position where its x-arm (y-arm) points North-East (North-West), by the angle indicated in the legend next to the label ``1T+1L-''. The triangular detector in the EMR region is always in its default position. The CE detector is rotated clockwise, starting from its default position, by the angle indicated in the legend next to the label ``CE-''. The definition and calculation of the parameters as well as the generation and selection of the BNS events are discussed in the text.}
        \label{figCBCOrientationStudy_1T1LCE_bns}
\end{figure*}

We have also studied the impact of the relative orientation of the detectors in the considered ET configurations. \\
The 1T displays a performance that is independent of its orientation for symmetry reasons. \\
The results obtained for the 2L by rotating clockwise the L-shaped detector in Sardinia with respect to the one in the EMR region by an angle comprised between  $0^\circ$ and $45^\circ$ are presented in Figs~\ref{fig:bbh_orientation} and~\ref{fig:bns_orientation}. As already noted in~\cite{COBA_Branchesi:2023mws}, the performance of two aligned detectors ($0 ^\circ$ curves) in terms of sky localisation and luminosity distance is significantly degraded. However, the performance obtained with an angle of $30^\circ$ between the two L-shaped detectors is only marginally degraded compared to $45^\circ$, with the largest reduction, by a factor of about 1.5, observed in the inclination angle curve in the case of BNS events\footnote{Moreover, a prior study~\cite{Begnoni:2025oyd} found that when requirements are imposed simultaneously on the uncertainty in multiple parameters the difference between the CBC performance of different orientations is further reduced and suggested that additional scientific cases, like SGWB, should be used to inform the choice of an optimal orientation. An optimisation based on CBC and SGWB was proposed in~\cite{3G_network_Ebersold:2024hgp}. This latter study considered explicitly, with a different metric for CBC performance, a network that included two L-shaped CE detectors and concluded that a balanced configuration for the two ET L-shaped detectors would have a tilt of about $26^\circ$ ($180^\circ+26^\circ$) between the x-arm (y-arm) of the detector in Sardinia and the y-arm (x-arm) of the detector in the EMR region.}   
This result is particularly interesting in light of the results obtained in Sec.~\ref{sec:SGWB_study}, where angles between the two L-shaped detectors greater than about $30^\circ$  significantly penalise the SGWB scientific case.
The results obtained for two ET L-shaped detectors with the addition of CE at different angles are also presented in Figs.~\ref{fig:bbh_orientation} and~\ref{fig:bns_orientation}. They have been obtained for two different angles between the ET L-shaped detectors, $45^\circ$ and $30^\circ$, while CE is rotated clockwise starting from its default orientation. It is evident that the addition of CE to the 2L network reduces significantly the dependence of the CBC performance on the relative orientation of the 2L and CE detectors. \\
Different orientations of the L with respect to the T in the 1T+1L, with or without CE, yield no statistically significant differences in the CBC event distributions, as can be appreciated in Figs~\ref{figCBCOrientationStudy_1T1LCE_bbh} and ~\ref{figCBCOrientationStudy_1T1LCE_bns}.
This feature of the 1T and 1T+1L is attractive because, contrary to the 2L, any technical construction constraint on the relative orientation of the detectors will not impact their CBC performance.


\section{Stochastic gravitational-wave background}
\label{sec:SGWB_study}
\subsection{Method}
An SGWB signal is expected to arise from the superposition of the GW signals that are either too weak or too numerous to be individually detected \cite{Allen:1997ad, Romano:2016dpx}. 
An SGWB signal can be either of astrophysical \cite{astro_SGWB_Regimbau:2022mdu} or cosmological origin \cite{Maggiore:1999vm, cosmo_SGWB_Caprini:2018mtu}. The latter case can result from early-universe processes like inflation, reheating, phase transitions, cosmic strings, or primordial black holes.

In this section, we present results on  the sensitivity of the three ET configurations, with and without additional detectors, to isotropic SGWB signals. The sensitivity is evaluated by employing the \textit{power-law integrated} (PI) \textit{curve} formalism introduced in \cite{PI_curves_Thrane:2013oya}. The formalism is briefly outlined below.

Under the assumption of Gaussianity, an isotropic and unpolarised SGWB can be fully characterised by 
the (1-sided) GW strain power spectral density $S_{h}(f)$. This latter quantity is related to the dimensionless GW energy density fraction $\Omega_{\rm gw}(f)$ by the following relation 
\begin{equation} 
\label{eq:Om_to_Sh}
S_{h}(f) = \frac{3 H_0^2}{2\pi ^2}\frac{\Omega _\mathrm{gw}(f)}{f^3} \equiv S_{0}(f) \Omega_{\rm gw}(f).
\end{equation}
The SGWB signal is usually assumed to follow a simple power law in frequency characterised by a spectral index $\alpha$
\begin{equation}
\label{Omegabetafref}
\Omega_{\mathrm{gw}}(f)=\Omega_{\alpha}\left(\frac{f}{f_{\mathrm{ref}}}\right)^{\alpha},  
\end{equation}
where $f_{\rm ref}$ is an arbitrary reference frequency. 

Searches for SGWB signals with ground-based detectors cross-correlate data from pairs of detectors. The baseline distances among different detector pairs and the relative detectors' orientations play a key role in the sensitivity of the detector network to GWBs. 
It can be proven~\cite{Allen:1997ad, PI_curves_Thrane:2013oya} that the expectation value for the SNR in a cross-correlation search for an unpolarized and isotropic SGWB signal is given by\\
\begin{equation}
    \label{eq:PI_curves_SNR}
    \SNR = \sqrt{2T\delta f} \sqrt{N_{\rm bins}} \expval{\frac{S_{h}^{2}}{S_{\rm eff}^{2}}}^{1/2},
\end{equation}
\\
where the angle brackets denote the average\footnote{Explicitly, $\expval{x} = \frac{1}{\Delta f}\int_{f_{\rm min}}^{f_{\rm max}} x(f) \,\dd f$.} over the frequency band $\Delta f \equiv N_{\rm bins}\, \delta f$ considered in the search, $T$ is the coincident observation time, $\delta f$ the frequency bin width, $N_{\rm bins}$ the number of frequency bins, and $S_{\rm eff}(f)$ the effective strain noise power spectral density
for the detector network defined by
\begin{equation}
    \label{eq:PI_curves_S_eff}
    S_{\mathrm{eff}}(f) \equiv \left(\sum_{I=1}^{N_{\rm det}}\sum_{J>I}^{N_{\rm det}} \frac{\Gamma_{IJ}^{2}(f)}{P_{n_{I}}(f)\, P_{n_{J}}(f)}\right)^{-1/2}.
\end{equation}
In this last expression the summations are over the $N_{\rm det}$ detectors in the network, $P_{n_{I}}$ is the noise power spectral density of detector $I$ in the network and $\Gamma_{IJ}(f)$ are the overlap reduction functions (ORFs) \cite{ORF_Christensen:1992wi, ORF_Flanagan:1993ix} that encode the geometry of the detector network.

By using equations \eqref{eq:Om_to_Sh} and \eqref{Omegabetafref}, one can invert equation \eqref{eq:PI_curves_SNR} to obtain $\Omega_{\alpha}$ and write the PI curve as\footnote{One can also derive the properties of the PI curve analytically, see \cite{PI_curve_Belgacem:2025oom}.}
\begin{equation}
\label{eq:PI_curve_defintion}
    \Omega_{\mathrm{PI}}(f) = \max_{\alpha} \left\{\Omega_{\alpha} \left(\frac{f}{f_{\mathrm{ref}}}\right)^{\alpha}\right\}.
    \end{equation}
The PI curve constructed in this way has the property that if a predicted power-law SGWB signal curve lies below (above, in some frequency range) the PI curve, then it will be observed with an SNR expectation value smaller (greater) than the SNR expectation value used to construct the PI curve.  

\subsection{Results}
\begin{figure*}
    \centering
    \includegraphics[width=0.49 \textwidth]{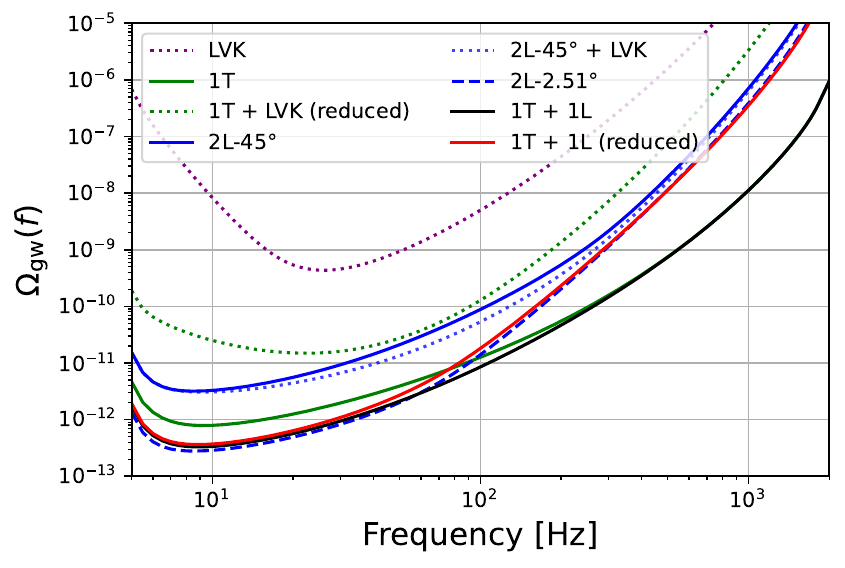}
    \includegraphics[width=0.49 \textwidth]{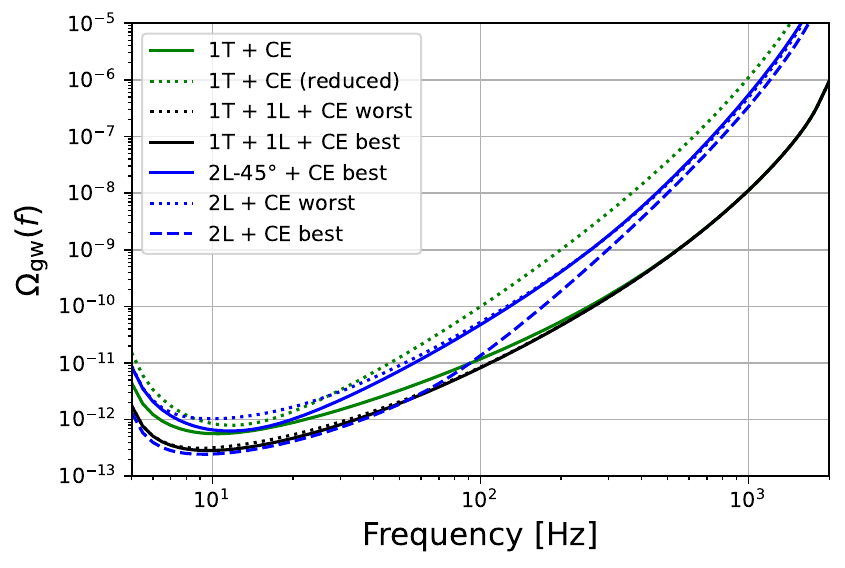}
    \caption{Left panel: PI curves, as defined in the text, for different configurations of an ET-only network, with or without the LVK detector network at design sensitivity. 
    The angle between the two L-shaped detector is indicated after ``2L-'' in the legend. Maximum alignment for this option corresponds to $2.51^\circ$.
    The 1T+1L \textit{reduced} configuration only uses cross-correlation data from the three pairs of distant detectors, as discussed in the text. Right panel: PI curves for different ET configurations when including an additional CE detector in the network. The ``best'' (``worst'') labels correspond to the configurations providing the lowest (highest) minimum of the PI curve in Fig.~\ref{fig:PI_curve_min}.
    }
    \label{fig:PI_curves}
\end{figure*}

\begin{figure*}
    \centering
    \includegraphics[width=0.49\textwidth]{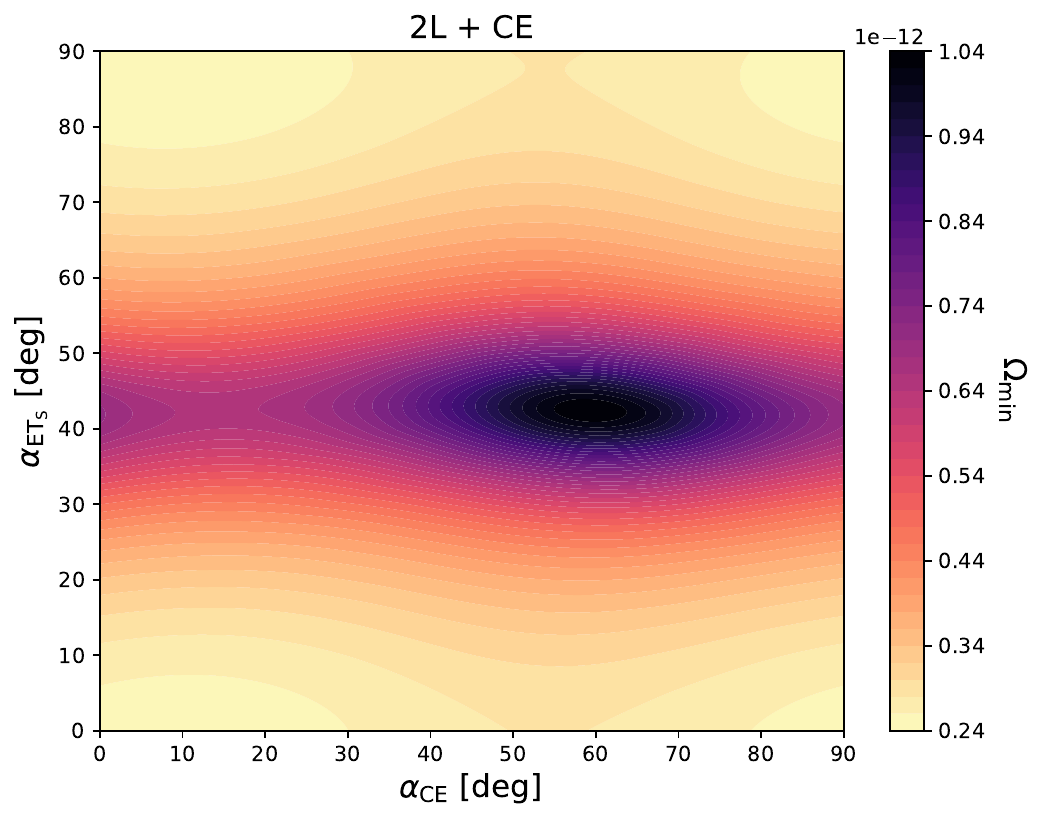}
    \includegraphics[width=0.49\textwidth]{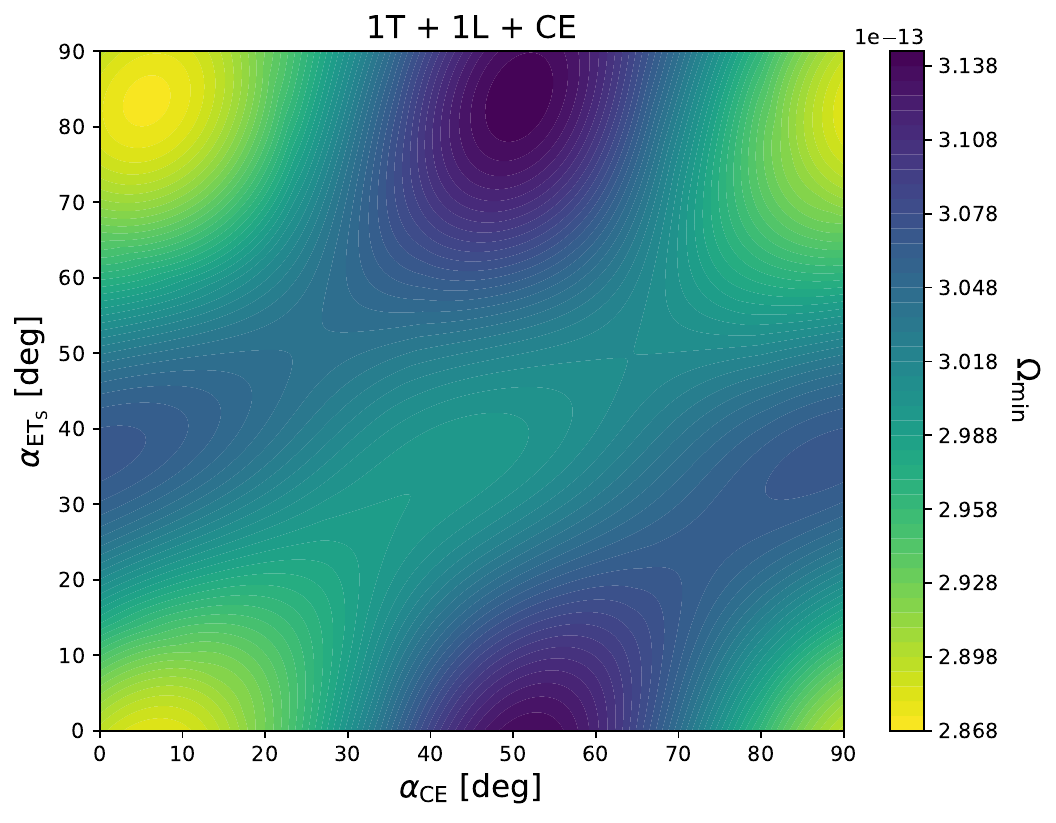}
    \includegraphics[width=0.49\textwidth]{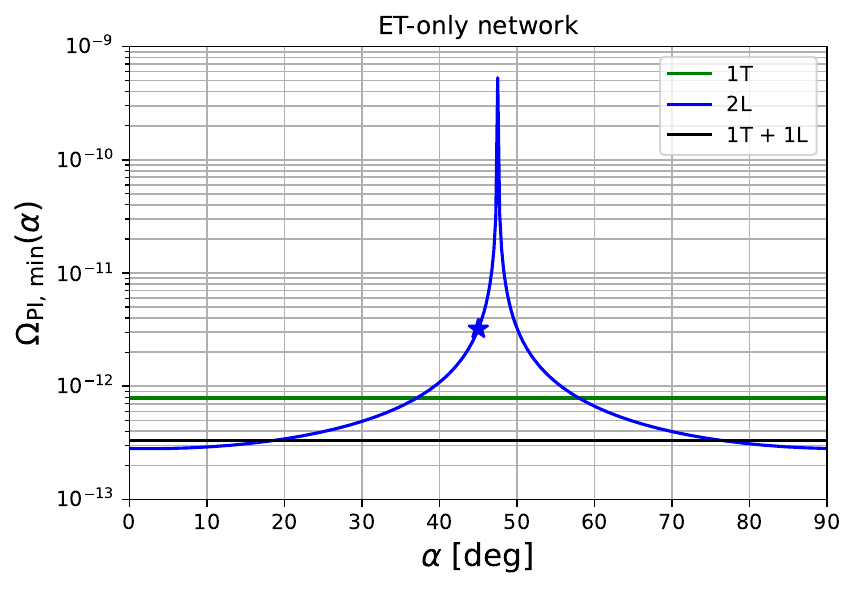}
    \caption{Minimum of PI sensitivity curve for different orientations of the ET and CE detectors. Top left panel: The ET L-shaped detector in the EMR region is kept fixed with the x-arm (y-arm) pointing the local East (North). The ET L-shaped detector in Sardinia and the CE in Idaho are rotated counterclockwise by angles $\alpha_{\rm ET_{S}}$ and $\alpha_{\rm CE}$, respectively, starting from the same local position as the ET L-shaped detector in the EMR region. The angles are in the $[0°, 90°]$ interval, with a spacing of $0.25°$. Top right panel: Same as in the top left panel, but with the ET L-shaped detector in EMR replaced by a triangle with a side parallel to the West-East local direction and the opposite vertex pointing North. Bottom panel: the three ET stand-alone ET configurations. Starting from the same initial positions defined above, the triangle for the 1T and 
    the ET L-shaped detector in Sardinia, for the 2L and 1T+1L, are rotated clockwise by an angle $\alpha$. 
    The blue star corresponds to the 2L configuration with an angle of $45^\circ$ between the two detectors.}
    \label{fig:PI_curve_min}
\end{figure*}

We evaluate the PI curve for the different detector configurations and networks using equation \eqref{eq:PI_curve_defintion}.
We assume a nominal $T_{\rm obs} = 1$ yr, an SNR = 1, and a frequency band between 5 and 2000 Hz with a 0.5 Hz binning. We present two sets of results: full PI curves as a function of frequency for given detector configurations in Fig.~\ref{fig:PI_curves}, and the minimum of the PI curve as a function of the relative orientation among the detectors in a network in Fig.~\ref{fig:PI_curve_min}. By construction, all these results are obtained under the assumption of Gaussian, non-correlated noise across detectors.

The results in the left panel of Fig.~\ref{fig:PI_curves} show that 1T+1L achieves a sensitivity that is between one and two orders of magnitude better than 2L, depending on the frequency. In fact, as already noted in previous studies~\cite{COBA_Branchesi:2023mws, 3G_network_Ebersold:2024hgp, 3G_network_Maggiore:2024cwf, 3G_network_Iacovelli:2026ixl, orintation_jishnu:Christensen:2026yyl}, any tilt angle between the two L-shaped detectors close to $45^{\circ}$ strongly penalises the SGWB scientific case. Two aligned L-shaped detectors, corresponding to a tilt angle of $2.51 ^\circ$ at the considered detector locations, 
would instead provide similar sensitivity as 1T+1L at low frequency. Above 100 Hz, the performance of 2 aligned L-shaped detectors is anyway strongly degraded as a result of the ORF suppression factor 
that reflects the $\mathcal{O}$(1000 km) distance between the detectors.
The 1T+1L performs about 2 to 3 times better than 1T up to about 100 Hz. Beyond this frequency the performance of these two configurations is similar as a result of the reduced impact of the distant detector. 

As mentioned above, these results are obtained under the assumption of absence of correlated noise, which could be problematic for SGWB searches~\cite{Janssens:2022xmo, Janssens:2021cta, Janssens:2022tdj} and could appear in co-located detectors. Methods based on quantifying and subtracting~\cite{Thrane:2014yza} or modelling correlated noise~\cite{Caporali:2025mum} have been proposed to avoid that correlated noise introduces a bias in searches for SGWBs. 
In this paper, instead, we exploit the possibility available in 1T+1L 
of using for the SGWB searches only cross-correlation data from the three pairs of distant detectors. Each pair is composed of one V of the triangle and the distant L. 
This possibility is of course also available in any configuration comprising 1T and additional distant detectors.
The expectation value of the final estimator will not be affected by the possible presence of correlated noise in the data streams of the V-shaped detectors. This feature is important because it implies that the possible presence of correlated noise will not mimic an SGWB signal. On the other hand, the standard deviation of the estimator, which affects the sensitivity, will change and in the extreme, most pessimistic case in which the noise in the 3 detectors is equal and fully correlated and the three V-shaped detectors are all aligned, it will only be a factor of $\sqrt{3}$ larger than if there was no correlation.
The sensitivity of this ``reduced'' configuration in the absence of correlated noise
is almost the same as that of the full 1T+1L below 100 Hz and close to the one provided by 2 aligned L-shaped detectors at higher frequency.  

The inclusion of LVK does not bring significant improvements to any of the ET configurations. This conclusion can be appreciated from the 2L+LVK curve reported in the left panel of Fig.~\ref{fig:PI_curves}. The LVK network would however enable the ``reduced'' option for the 1T. Nevertheless, the resulting performance, which is also shown in the left panel of Fig.~\ref{fig:PI_curves}, is worse than for 2L, especially at low frequency.   

As it can be seen in the right panel of Fig.~\ref{fig:PI_curves}, the inclusion of CE in the network improves significantly the performance of 2L below 100 Hz. The 1T option is also improved in the same frequency range, but by a smaller factor compared to 2L. Instead, CE only slightly improves the SGWB performance of 1T+1L. 
Even with CE in the network, the 1T+1L performance remains about a factor of 2 better than both 1T and 2L (with an angle of $45^\circ$) below 100 Hz, while at higher frequency it would perform one order of magnitude (or more) better than 2L, and approximately the same as 1T. Like LVK, CE would also enable the ``reduced'' option for the 1T, but with a much better result, especially at low frequency.

Another attractive feature of 1T+1L is an SGWB performance that does not appreciably depend on the relative orientation of the L-shaped and triangle detectors. The same holds for 1T whose performance does not depend on the orientation of the triangle. Instead, a configuration with 2 L-shaped detectors displays a strong dependence on the relative orientation of the detectors, which will possibly have to cope with construction constraints. These dependencies are displayed in the bottom panel of Fig.~\ref{fig:PI_curve_min} that reports how the minimum of the PI curve varies as a function of the relative orientation between the detectors for the three considered ET configurations.  

The dependence of the minimum of the PI curves on the relative orientation of the ET and CE detectors is displayed instead in the top two panels of Fig.~\ref{fig:PI_curve_min}.  
In these plots, the minimum of the PI curve varies when the ET L-shaped detector placed in Sardinia and the CE detector are rotated counter-clockwise with respect to the ET detector (triangle or L-shaped) placed in EMR by the angles $\alpha_{\rm ET_{S}}$ and $\alpha_{\rm CE}$, respectively. One can see that the 1T+1L performance only varies by at most 10\% over the entire space of parameters, while this variation is almost one order of magnitude for an ET configuration composed of two L-shaped detectors. The combination of angles yielding the lowest (highest) value of the minimum of the PI curve, is used to draw the PI curves labeled ``best'' (``worse'') in the right panel of Fig.~\ref{fig:PI_curves}.

When combining the results obtained in this section with those of Sec.~\ref{sec:CBC_study} it can also be concluded that a relative orientation of $30^\circ$ between the two L-shaped detectors would provide a more balanced performance between the CBC and SGWB scientific cases for this option.

\section{Bursts}
\label{sec:Bursts}
The detection of unmodeled, short-duration (order of 1 second or less) GW signals, also called {\it bursts}, represents an important part of the ET physics program. Bursts are expected to be produced as a result of phenomena like core-collapse supernovae~\cite{Janka:2012wk}, magnetars~\cite{10.1093/mnras/staa2322} or cosmic strings~\cite{PhysRevD.64.064008}. Longer duration bursts can also come from accretion of matter from newly-born NSs or BHs~\cite{1995ApJ...442..259L}.

The lack of a waveform to be compared to the observed data implies that the key ingredient for a first observation of GW bursts is keeping the false alarm rate (FAR) as low as possible. In this respect the coincident detection with as many detectors as possible and the possibility to efficiently and confidently reject spurious coincidences of glitches, or Gaussian noise fluctuations, are invaluable handles. The sensitivity of a GW detector to a burst signal, which is driven by the SNR, is of course also important for discovery, but from the results obtained in Sec.~\ref{sec:CBC_study}, it is reasonable to expect for short-duration bursts small differences between 1T+1L and 2L, as well as values 30-50\% smaller for 1T. 

Experience from running and past detectors~\cite{LIGO:2024kkz, Virgo:2022ysc, Zevin:2016qwy} is extremely precious to assess the rate and nature of glitches, but unfortunately the origin, whether instrumental or environmental\footnote{Environmental disturbances are susceptible to produce correlated glitches in co-located detectors, whereas purely instrumental ones would only affect the concerned detector.}, of many glitch classes remains unknown. As far as co-located detectors are concerned, LIGO operated in the past the two co-located H1 and H2 detectors~\cite{Blackburn_2008, Abbott_2007}, but it should be underlined that some of the main mirrors of these detectors were only a few meters apart and the two detectors were sharing the same beam tube~\cite{Coyne2008}. The ET detectors of the triangle will instead feature~\cite{ET_Design_2025} separate beam tubes and a minimal separation between main mirrors of a few hundred meters. Such distances could not only reduce very significantly the rate of correlated glitches, but also facilitate the identification of the glitch source in those cases where the propagation of the disturbance at the origin of the glitches is of mechanical nature (e.g. seismic waves), thanks to the observable time delay between the signals observed in different detectors. In any case ET, especially in the low-frequency range, will explore an uncharted territory and therefore uncertainty remains in the rate and nature of glitches appearing in its co-located detectors.

In principle, an ideal, minimal configuration to minimise the FAR for burst signals would be represented by two perfectly aligned and co-located detectors. In this case a genuine GW burst signal would be observed at the same exact time and with the same exact shape and strain amplitude in both detectors, leaving the smallest possible probability of occurrence to spurious coincidences of uncorrelated glitches in the two detectors. Distant, aligned detectors would still record the same shape and strain amplitude for the signal, but at different times depending on the source sky position. The FAR would grow linearly with the distance between the detectors~\footnote{The FAR linear growth is in order only for distances between the two detectors greater than about $c\, \delta t$, where $c$ is the speed of light and $\delta t$ the detector signal arrival time resolution. This distance is about 300 km for $\delta t = 1$ ms, for instance. Below this distance the FAR is to first approximation constant.}(see e.g.~\cite{Schutz1989DataAnalysisNetworks}). In the case of misaligned, distant detectors, like the 2L configuration considered in this study, the GW signal projects differently at the output of the two detectors, depending on signal polarisation and sky position of the source. This configuration will therefore feature higher FAR compared to two aligned L-shaped detectors, given that the two signal detector measurements and the time delay between the two measurements are not sufficient to fully constrain GW signals. 

The 1T configuration provides three co-located (not aligned) detectors, and comes with the null stream feature, which is extraordinarily attractive for FAR reduction because genuine GW signals would cancel out in the null stream, contrary to what would normally happen to coincident glitches\footnote{Accidental (quasi-)cancellation of coincident glitches in the null stream can however still happen. The glitch residual in the null stream will have to be identified in the fluctuating sum of the detectors' Gaussian noise.}. Glitches of environmental nature could however increase the rate of candidate signals appearing simultaneously in the 1T co-located detectors.

The inclusion of additional detectors, like LVK, CE, or an extra ET-like L-shaped detector as in the 1T+1L, would be extremely beneficial as it would enable coincidences with a larger number of distant detectors and thus reduce dramatically the FAR~\footnote{The SNR of the LVK configuration considered in this paper is about a factor of 10 lower than ET. Therefore for this specific LVK configuration to be useful in burst searches, the considered signals must be sufficiently loud.}.\\
In particular, the 1T+1L configuration would come with the possibility of 4-detector coincidences, all the advantages of the triangle, null stream and simultaneity of genuine GW signals, and the invaluable information from the distant L-shaped detector. In stand-alone ET operation, spurious glitch coincidences would be maximally rejected and genuine GW signals would be maximally constrained and better measured thanks to information from 4 detectors and 1 accurately measured time delay, as opposed to 2 detector measurements and 1 time delay for 2L and 3 detector measurements for 1T.

A dedicated study, which goes beyond the scope of this document, needs to be conducted in order to accurately quantify the performance of the various ET configurations, with or without additional detectors and under different realistic scenarios for detector glitches.

\section{Continuous waves}
\label{sec:CW}
CWs correspond to long-duration (observable over up to several years), quasi-monochromatic GW signals. They are expected to be produced by a number of phenomena: isolated, rotating, non-axisymmetric NSs, opening the way to study the NS internal dynamics~\cite{Lasky_2015, Glampedakis:2017nqy}; the possible accumulation around spinning BHs of yet-unobserved elementary boson particles~\cite{Brito:2017zvb}, which could be dark matter candidates~\cite{ARBEY2021103865}; the direct interaction of dark matter particles with the GW detector mirrors~\cite{Pierce:2018xmy, Miller:2020vsl}; or the early inspiral of binary systems composed of sub-solar mass primordial BHs~\cite{Miller:2020kmv}.

Like for unmodelled burst and SGWB signals, no CW signal has been detected to date with ground-based GW detectors. For a first detection, the sensitivity of a GW detector to a CW signal is the main relevant figure of merit to consider. Because of their long duration and the large number of parameters in the exact signal model, CW searches can be extremely challenging from a computational point of view. For this reason their sensitivity strongly depends~\cite{Riles_2023} on the adopted analysis method, whether coherent or semi-coherent, and the explored signal source parameter space, which depends on the prior information assumed on the source.

Following the approach of recent works~\cite{Pagliaro_2023, 10.1093/mnras/stac3665, Tenorio:2023hzh, 10.1093/mnras/staf774}, it is possible to compare the sensitivity of different detector configurations to a given CW search by introducing the sensitivity depth~\cite{PhysRevD.91.064007} of the search:
\begin{equation}
\label{eq:sensitivitydepth}
\mathcal{D}(f)=\frac{h_0(f)}{\sqrt{S_n(f)}}.
\end{equation}
where $S_n(f)$ is the noise power spectral density of the detectors in the configuration, which is a function of the frequency $f$, and $h_0(f)$ is the sensitivity of the search that can be defined as the upper limit~\footnote{Averaged over all sky positions, polarisations, and inclinations of the source.}, at a given confidence level, on the amplitude of a CW signal of frequency $f$ achievable by the search with a given observation time.\\
To first approximation, the sensitivity depth does not change when computed on configurations with the same number and shape of detectors. Therefore, by taking the sensitivity depth of a CW search that has already been performed with a given detector configuration (e.g. $\mathcal{D}_{\rm LIGO}$ in a search performed with the two LIGO detectors), the sensitivity of the same search on a data sample of the same duration and collected with the 2L ET configuration can be obtained from equation~\eqref{eq:sensitivitydepth} by simply using the ET noise power spectral density corresponding to 15 km long arms, $\sqrt{S_n^{\rm L15}}$. 
The sensitivity of the ET 1T configuration can then be obtained by applying proper correction factors. The latter factors account for the different number of identical detectors in the 2L and 1T ($\sqrt{2/3}$, because the SNR is proportional to the square root of the number of detectors~\cite{PhysRevD.72.063006}), the angle $\alpha$ between the detector arms ($1/\sin{\alpha}=2/\sqrt{3}$, because the detector antenna pattern functions, and thus the SNR~\cite{PhysRevD.58.063001}, are proportional to the sine of this angle), and for differences in the detectors arm length between the 2L and 1T ($3/2$, because to first approximation the displacement noise is inversely proportional to the arm length\footnote{For the sake of simplicity, this factor is applied in this context to the amplitude spectral density over the full frequency range, which is of course a coarse approximation~\cite{LIGOScientific:2016wof}.}). When these factors are applied, one gets:
\begin{equation}
\label{eq:CWsensitivity}
\frac{\sqrt{S_n^{\rm L15}}}{\mathcal{D}_{\rm LIGO}}=h_0(2L)=h_0(1T+1L)=\frac{1}{\sqrt{2}} h_0(1T) 
\end{equation} 
where the result for 1T+1L has been obtained by combining two detectors, 1L with 15 km arms and 1T, which turn out to have the same CW sensitivity when appropriate correction factors are applied. \\ 
There are therefore small differences in sensitivity to be expected in CW searches between 1T+1L and 2L, while 1T would be some 30\% less sensitive. 

Nevertheless a number of points need to be noted. First, the actual search sensitivity will also depend on non-Gaussian noise features, including possible correlations across detectors. Second, the availability of the null stream in the configurations that include a triangle will be a very powerful tool in following-up candidates passing the threshold for detection. Third, the availability of more detectors could be an advantage in those searches that first identify candidates in individual detectors and then require
coincidences across different detectors to reduce to a computationally reasonable level the number of CW signal candidates to follow-up. It has been shown that in a semi-coherent, all-sky search~\cite{PhysRevD.90.042002} with two detectors and under the assumption of purely Gaussian noise, starting from more than $10^{17}$ points in the explored parameter space, it is possible to go from $10^9$ candidates per detector to only $O(10)$ candidates to follow-up.
In this perspective, triple coincidences are possible with the 1T configuration, while 1T+1L provides the possibility of quadruple coincidences. Thus, if the number of candidates to follow-up is fixed, then requiring a larger number of coincidences allows in principle to lower the thresholds for selecting CW candidates in the dataset of each individual detector, which in turn means increasing the sensitivity of the search.\\ 
It is also important to underline that even for fully coherent CW searches the number of required match filtering templates does not scale with the number of detectors and the required computing resources to perform a CW analysis will only increase linearly with this number~\cite{PhysRevD.75.023004}. \\
Finally, in light of the above discussion, it can also be expected that the addition of a network like LVK is not expected to bring any visible improvement to CW searches performed with any of the ET configurations, while the opposite is true for the addition of CE.

A dedicated study, which goes beyond the scope of this document, needs to be performed to assess the actual impact of non-Gaussian detector noise, the null stream and multiple detectors.

\section{Conclusions}
We have compared the scientific performance of the two baseline ET configurations that are currently under consideration, a single triangle with a 10 km long side (1T) and 2 distant L-shaped detectors tilted by about 45$^\circ$ with 15 km long arms (2L), in both stand-alone mode and in the presence of additional detectors, either a single L-shaped CE detector with 40 km long arms, or the LVK network with the sensitivity expected for the O5 run. 
We have also considered an alternative, hypothetical ET configuration consisting of one 10 km triangle and one 15 km L-shaped detector (1T+1L). This configuration is not currently under consideration and would clearly require additional resources for construction, maintenance and upgrade. We have obtained results in simplified setups for the CBC and SGWB scientific cases, while searches for burst and continuous wave signals have been simply discussed with basic principles.
We find that the 1T+1L performs significantly better than 1T and 2L in both the CBC and SGWB scientific cases. In addition, the 1T and 1T+1L would feature all the advantages of the null stream provided by the triangular detector.  
The availability of the null stream and multiple detectors, as in the 1T and 1T+1L configurations, is also expected to significantly boost the performance of searches for new signals, like bursts, by reducing the false alarm rate. At the same time, as already pointed out in previous studies, the co-located detectors of a triangular detector can suffer from the possible presence of correlated noise of environmental origin, which could instead increase the false alarm rate. 
The impact of the null stream, false alarm rate reduction with multiple detectors, non-Gaussian noise and correlated noise, could not be incorporated nor quantified in the results presented in this document and should be the subject of future dedicated studies.

We have also studied the impact of the relative orientation of the ET detectors in the multi-site configurations and find that the 1T+1L configuration, contrary to an option with two L-shaped detectors, does not display any significant dependence on the relative orientation of its two distant detectors, thus being basically unaffected by possible technical construction constraints, like the 1T. We also find that a better compromise in terms of balanced scientific performance for an option with two L-shaped detectors could be a relative orientation between the two detectors of about $30^\circ$.

The addition of CE improves significantly the scientific performance of the network reducing at the same time the differences in performance between the various ET configurations. CE also reduces the impact of the relative orientation between detectors in the ET option with 2 L-shaped detectors. 
In terms of sky localisation of CBC signals, even the inclusion of LVK to the network improves significantly the performance of the ET configurations and reduces their differences, though to a lesser extent than the inclusion of CE. 
Finally, we have shown that the impact of possible correlated noise in co-located detectors on the SGWB performance can be addressed with the addition of extra detectors. This addition would allow cross-correlation data from co-located detectors to be disregarded. For both 1T+CE and 1T+1L the performance degradation is only significant above 100 Hz and simply results from the distance between the considered detectors.

\section{Acknowledgements}

We are grateful to the authors of~\cite{Iacovelli:2022bbs,GWFAST_Iacovelli:2022mbg} and~\cite{COBA_Branchesi:2023mws} for making the \verb|GWFAST| code and the CBC datasets, respectively, freely available. We thank Aaron Goodwin-Jones, Justin Janquart, David Keitel and Jo van den Brand for the useful discussions. 
Antonin Oswald is supported by the Actions de Recherche Concertées (ARC) of the Fédération Wallonie-Bruxelles, under grant ``GRAVIT-AI''. 
Computational resources have been provided by the supercomputing facilities of the University of Louvain (CISM/UCL) and the Consortium des Équipements de Calcul Intensif en Fédération Wallonie Bruxelles (CÉCI) funded by the Fond de la Recherche Scientifique de Belgique (F.R.S.-FNRS) under convention 2.5020.11 and by the Walloon Region.

\bibliography{bib}

\end{document}